\documentclass[aps,prd,twocolumn,superscriptaddress,nofootinbib,floatfix,noshowpacs]{revtex4-1}
\pdfoutput=1
\newif\ifcomment
\usepackage{amsmath,amssymb}
\usepackage{color,hyperref,url}
\usepackage{listings}
\usepackage{slashed}
\usepackage[pdftex]{graphicx}
\usepackage{epstopdf}
\usepackage{epsfig}
\usepackage{grffile}
\usepackage{relsize}
\graphicspath{{./img/}}
\usepackage{soul}

\newcommand{\al}{\alpha}
\newcommand{\be}{\beta}

\newcommand{\beq}{\begin{equation}}
\newcommand{\eeq}{\end{equation}}
\newcommand{\ba}{\begin{array}}
\newcommand{\ea}{\end{array}}
\newcommand{\bea}{\begin{align}}
\newcommand{\eea}{\end{align}}
\newcommand{\bi}{\begin{itemize}}
\newcommand{\ei}{\end{itemize}}
\newcommand{\ben}{\begin{enumerate}}
\newcommand{\een}{\end{enumerate}}
\newcommand{\bc}{\begin{center}}
\newcommand{\ec}{\end{center}}
\newcommand{\bl}{\begin{flushleft}}
\newcommand{\el}{\end{flushleft}}
\newcommand{\br}{\begin{flushright}}
\newcommand{\er}{\end{flushright}}

\newcommand{\nn}{\nonumber \\}
\newcommand\Eqn[1]{Eq.~(\ref{#1})}  
\newcommand\Fig[1]{Fig.~\ref{#1}} 

\newcommand{\fm}{{\rm fm}}

\newcommand{\eV}{{\rm eV}}

\newcommand{\GeV}{{\rm GeV}}

\begin{document}
\title{The chiral anomaly and the pion transition form factor: beyond the cutoff}
\author{Hao Dang}%
\email{haodang@mail.nankai.edu.cn}
\affiliation{School of Physics, Nankai University, Tianjin 300071, China}
\author{Zanbin Xing}%
\email{xingzb@mail.nankai.edu.cn}
\affiliation{School of Physics, Nankai University, Tianjin 300071, China}
\author{M. Atif Sultan}%
\email{atifsultan.chep@pu.edu.pk}
\affiliation{School of Physics, Nankai University, Tianjin 300071, China}
\affiliation{  Centre  For  High  Energy  Physics,  University  of  the  Punjab,  Lahore  (54590),  Pakistan}
\author{Kh\'epani Raya}
\email{khepani.raya@dci.uhu.es}
\affiliation{Department of Integrated Sciences and Center for Advanced Studies in Physics, Mathematics and Computation, University of Huelva, E-21071 Huelva, Spain.}
\author{Lei Chang}%
\email{leichang@nankai.edu.cn}
\affiliation{School of Physics, Nankai University, Tianjin 300071, China}
\date{\today}
\begin{abstract}
In the presence of a momentum cutoff, effective theories seem unable to faithfully reproduce the so called chiral anomaly in the Standard Model. A novel prospect to overcome this related issue is discussed herein via the calculation of the $\gamma^{*}\pi^0\gamma$ transition form factor, $G^{\gamma^* \pi^0 \gamma}(Q^2)$, whose normalization is intimately connected with the chiral anomaly and dynamical chiral symmetry breaking (DCSB). To compute such transition, we employ contact interaction model of Quantum Chromodynamics (QCD) under a modified rainbow ladder truncation, which automatically generates a quark anomalous magnetic moment term, weighted by a strenght parameter $\xi$. This term, whose origin is also connected with DCSB, is interpreted as an additional interaction that mimics the complex dynamics beyond the cutoff. By fixing $\xi$ to produce the value of $G^{\gamma^* \pi^0 \gamma}(0)$ dictated by the chiral anomaly, the computed transition form factor, as well as the interaction radius and neutral pion decay width, turn out to be comparable with QCD-based studies and experimental data.
\end{abstract}

\maketitle
\section{introduction}\label{sec:int}
Conserved currents in classical theory may be violated by quantum corrections, an outcome referred to as anomaly. One of the most notable anomalies in the Standard Model is the chiral anomaly  discovered by Adler, Bell and Jackiw in 1969~\cite{Bell:1969ts,Adler:1969gk}, which is responsible for the neutral pion decay, thereby having a significant impact in the $\gamma^{*}\pi^0\gamma$ transition form factor (TFF). In addressing this transition process, there is a longstanding problem that in effective theories such as Nambu–Jona-Lasinio (NJL) like theories; namely, the anomalous neutral pion decay is smaller than the experimental value in the presence of a finite cutoff~\cite{Blin:1987hw,Alkofer:1992nh}, and the difference between theory and experiment relies on the regularization scheme. This problem is discussed extensively but there is not a completely satisfactory solution, see Refs.~\cite{Alkofer:1992nh,Bernard:1993wf,Blin:1987hw,RuizArriola:2002wr,Schuren:1993aj}. For instance, if the cutoff is removed, the chiral anomaly is faithfully reproduced\,\cite{Roberts:2010rn}. Nonetheless, one could argue that the cutoff itself is part of the effective theory and thus should not be neither changed nor removed in the calculation of observables. On the other hand, the role of the cutoff within a path integral derivation of the chiral
anomaly is clarified in Ref.~\cite{Alkofer:1992nh}, highlighting cutoff-dependent higher order contributions that are crucial for the chiral anomaly. In principle, it is impossible to explicitly calculate such contributions at all orders without making assumptions. However, as meticulously illustrated in Ref.~\cite{Lepage:1997cs}, once the cutoff is introduced in the effective theory, additional local interaction terms should be added to mimic the complex interaction triggered by higher order contributions that appear at shorter distances (or at higher energies) compared to the cutoff. These additional terms depend on the latter and must vanish upon its removal. This conception, however, has not been implemented so far in any effective theory study involving the chiral anomaly. In this work, we shall adopt these ideas in the calculation of the neutral pion decay and the corresponding $\gamma^{*}\pi^0\gamma$ TFF, by using an effective model within the framework of Dyson-Schwinger equations (DSEs), \emph{i.e.} the contact interaction (CI) model~\cite{Gutierrez-Guerrero:2010waf,Frederico:1992np}.

The DSE formalism has proven to be a powerful tool in studying the nonperturbative nature of Quantum chromodynamics (QCD) in the continuum~\cite{Roberts:1994dr,Fischer:2018sdj}, representing an ideal platform to study the static and structural properties of hadrons\,\cite{eichmann2016baryons,qin2019spectrum,cui2020kaon,arrington2021revealing,Roberts:2023lap}. Within this framework, the mass spectrum and structural properties of hadrons are governed by the relationship between the quark DSE and the bound-state Bethe-Salpeter (BS) and Feddeev equations \cite{SALPETER:1961cs,Faddeev:1961cs}. In fact, bound state equations would be related to the Green functions of the theory, in such a way that the resulting infinite system of integral equations must be truncated in a systematic way\,\cite{Chang:2009zb,Qin:2020jig}. A popular approach is the so called symmetry-preserving vector-vector contact interaction, originally introduced to study the properties of the pion~\cite{Gutierrez-Guerrero:2010waf,Roberts:2010rn}, in a relatively simple framework capable of preserving key features of QCD such as confinement and chiral symmetry
breaking. To date, the CI model has been employed to address numerous hadronic properties, including meson and baryon mass spectrum, various decay processes, form factors and parton distributions (see \emph{e.g} Refs.~\cite{Roberts:2010rn,Gutierrez-Guerrero:2010waf,Wilson:2011aa,Roberts:2011cf,Roberts:2011wy,Chen:2012qr,Segovia:2014aza,Serna:2017nlr,Zhang:2020ecj,Gutierrez-Guerrero:2021rsx,Raya:2021pyr,Cheng:2022jxe,Xing:2022mvk,Hernandez-Pinto:2023yin,Xing:2023eed,Zamora:2023fgl}); the emanating predictions, especially those concerning static properties, have provided valuable benchmark for both more sophisticated treatments of QCD and experiment.

To properly address meson properties, a consistent truncation of the DSE and BS equation (BSE) is crucial to preserve the symmetries of QCD. The traditional rainbow-ladder (RL) truncation is the common choice\,\cite{Maris:2002mz,Raya:2015gva,Chen:2016bpj,Ding:2019lwe}, among other, because it properly captures the Goldstone-boson nature of the pion\,~\cite{Munczek:1994zz,Bender:1996bb}. However, embedded within the CI, the RL truncation fails at reproducing the chiral anomaly due to the presence of cutoffs; only when those are removed it is possible to recover the anomaly~\cite{Roberts:2010rn}. A modified RL (MRL) truncation, recently proposed in Refs.~\cite{Xing:2021dwe,Wang:2022mrh}, would represent feasible an alternative. This truncation consistently generates the quark anomalous magnetic moment (AMM) term in the quark-photon vertex (QPV), while maintaining the relevant symmetries of QCD and leaving the pion static properties untouched. Following the spirit of Refs.~\cite{Alkofer:1992nh,Lepage:1997cs}, the AMM term can be interpreted as an additional interaction, between quark and photon, which mimics the dynamics beyond the cutoff. As required by this interpretation, we will also see in the following that the AMM term vanishes if one removes the CI-induced momentum cutoff under a proper regularization procedure.

The manuscript is organized as follows: in Sec.~\ref{sec:ci}, we briefly introduce the CI under the MRL truncation and expose how the structure of the QPV develops an AMM term. 
In Sec.~\ref{sec::ano}, we discuss the $\gamma^{*}\pi^0\gamma$ TFF in detail: in particular, its connection with the AMM term in detail, its connection with the chiral anomaly, and associated physical quantities. Conclusions and final remarks are presented in Sec.~\ref{sec::con}.

\section{contact interaction}\label{sec:ci}
A natural starting point for the calculation of the $\gamma^{*}\pi^0\gamma$ transition form factor is the quark DSE. This can be expressed mathematically as follows:
\footnote{We employ an Euclidean metric with $\{\gamma_\mu,\gamma_\nu\} = 2\delta_{\mu\nu}$; $\gamma_\mu^\dagger = \gamma_\mu$; $\gamma_5= \gamma_4\gamma_1\gamma_2\gamma_3$; and $a \cdot b = \sum_{i}^{4} a_i b_i$. The isospin symmetry is considered througout this work.}
\begin{equation}\label{eqn::Ngap}
    S^{-1}(p)=S_{0}^{-1}(p)+\frac{4}{3}g^2\int_q D_{\mu\nu}(p-q)\gamma_\mu S(q) \Gamma^{G}_\nu(q,p)\,.
\end{equation}
 where $\int_q\doteq\int\frac{d^4q}{(2\pi)^4}$ denotes a Poincar\'e invariant integration. The dressed quark propagator is fully characterized by two Dirac structures via
\begin{equation}\label{eq:quarkprop}
     S^{-1}(p) = Z^{-1}(p^2)(i\gamma \cdot p + M(p^2)\,,
\end{equation}
such that it maintains an analogy with its tree level counterpart, $S_0^{-1}(p)=i\gamma \cdot p + m$; here $m$ is the bare quark mass and  $M(p^2)$ denotes the so called mass function. The rest of the ingredients of Eq.\,\eqref{eqn::Ngap}, also known as gap equation, are defined as usual: $\Gamma^{G}_\nu$ and $g^2 D_{\mu\nu}$ represent, respectively, the fully-dressed quark-gluon vertex (QGV) and gluon propagator ($g$ is the Lagrangian coupling constant), each of which satisfy their own DSE. This interconnection yields an infinite number of coupled, nonlinear integral equations, which must be systematically truncated to study a physical system\,\cite{Qin:2020rad}. Typically, one assumes an appropriate form for the QGV that enable us to arrive at a tractable problem\,\cite{Sultan:2018tet,Serna:2018dwk}. In practice, this also requires replacing the gluon propagator by an effective one $g^2 D_{\mu\nu}(p-q)\rightarrow D_{\mu\nu}^{eff}(p-q)$. 
  
 In the CI model, the fully-dressed QGV is demoted to its tree level form, $\Gamma_\nu^G \to \gamma_\nu$, corresponding to the \emph{rainbow} approximation of the gap equation. The corresponding effective gluon propagator is defined as\,\cite{Gutierrez-Guerrero:2010waf}:
\begin{equation}\label{eqn::CI}
    g^2D^{eff}_{\mu\nu}(p-q)\rightarrow \frac{1}{m_G^2}\delta_{\mu\nu},
\end{equation}
where $m_G$ is a gluon mass-scale. Thus, the quark gap equation is expressed as:
\begin{equation}\label{eqn::gap}
    S^{-1}(p)=S_{0}^{-1}(p)+\frac{4}{3m_G^2}\int_q \gamma_\mu S(q) \gamma_\mu\,.
\end{equation}
This integral possess quadratic divergence that must be regularized in a Poincare invariant manner. The solution of Eq.\,\eqref{eqn::gap} then yields a rather simple form for the quark propagator,
\begin{equation}\label{eqn::quark}
    S^{-1}(p)=i \gamma \cdot p+M\,,
\end{equation}
where the mass function $M$ in Eq.\eqref{eq:quarkprop} becomes independent of the quark momentum $p$. Plugging \Eqn{eqn::quark} into \Eqn{eqn::gap}, one obtains the following nonlinear integral equation for $M$:
\begin{equation}\label{eqn::massf}
    M=m+\frac{16}{3m_G^2}\int_q\frac{M}{q^2+M^2}.
\end{equation}
Herein we adopt the symmetry preserving regularizaiton schemes described in Ref.~\cite{Xing:2022jtt}, which is based on the Schwinger's proper time method:
\begin{eqnarray}
I_{-2\alpha}(\mathcal{M}^2)&=&\int_{q}\frac{1}{(q^2+\mathcal{M}^2)^{\alpha+2}}\nn
                    &=&\int_{0}^{\infty}d\tau \frac{\tau^{\alpha-1}}{\Gamma(\alpha+2)}\frac{e^{-\tau\mathcal{M}^2}}{16\pi^2}\nn
\label{eqn::sreg}
&\rightarrow&\int_{\tau_{uv}^2}^{\tau_{ir}^2}d\tau \frac{\tau^{\alpha-1}}{\Gamma(\alpha+2)}\frac{e^{-\tau\mathcal{M}^2}}{16\pi^2}\nn
              I^{R}_{-2\alpha}(\mathcal{M}^2) &=&\frac{1}{16\pi^2}\frac{\Gamma[\alpha,\tau_{uv}^2\mathcal{M}^2]-\Gamma[\alpha,\tau_{ir}^2\mathcal{M}^2]}{\mathcal{M}^{2\alpha}\Gamma(\alpha+2)}\,,
\end{eqnarray}
 where $\tau_{uv}=1/\Lambda_{uv}$ and $\tau_{ir}=1/\Lambda_{ir}$ are ultraviolet (UV) and infrared (IR) regulators respectively; $\Lambda_{ir}\simeq \Lambda_{QCD}$ guarantees confinement by ensuring the absence of quark production thresholds, whereas $\Lambda_{uv}$ plays a dynamical role setting the scale of all dimensioned quantities. $\Gamma(n,z)$ is the incomplete gamma function. The label $R$ stands for regularized integrals and will be suppressed in the rest of the paper for simplicity. Thus, in terms of the so-called irreducible loop integrals (ILIs)~\cite{Xing:2022jtt}, \Eqn{eqn::massf} becomes
\begin{equation}\label{eqn::mass}
    M=m+\frac{16 M}{3m_G^2}I_2(\mathcal{M}^2)\,,
\end{equation}
such that $M$ is obtained by solving \Eqn{eqn::mass}.

Mesons are described by the bound state BSE, whose interaction kernel must be written in a self-consistent manner with the gap equation\,\,\cite{Chang:2009zb,Qin:2020jig}. In the MRL truncation, the meson BSE reads
\begin{eqnarray}\label{eqn::pion}
    \Gamma_{H}(P)=&-&\frac{4}{3m_G^2}\int_q \gamma_\alpha S(q)\Gamma_{H}(P)S(q-P)\gamma_\alpha\nn
    &+&\frac{4\xi}{3m_G^2}\int_q \tilde{\Gamma}_j S(q)\Gamma_{H}(P)S(q-P)\tilde{\Gamma}_j,
\end{eqnarray}
where $\Gamma_{H}(P)$ is the $H$-meson BS amplitude (BSA), with $P$ being the total momentum of the bound-state. Note the first line above, in conjunction with Eq.\,\eqref{eqn::gap}, define the RL truncation.
Conversely, the second line in \Eqn{eqn::pion} contains the Non-ladder (NL) pieces, $\tilde{\Gamma}_j=\left\{I_4,\gamma_5,\frac{i}{\sqrt{6}}\sigma_{\al\be}\right\}$. Finally,  $\xi$ is a strength parameter controlling the relative weight between the RL and NL contributions, such that $\xi=0$ recovers the traditional RL truncation.

The general form the pion BSA adopts within the CI-MRL truncation is:
\begin{equation}\label{eqn::pibsa}
    \Gamma_{\pi}(P)=i\gamma_5 E_\pi(P)+ \frac{\gamma_5 \gamma \cdot P}{M} F_\pi(P)\;,
\end{equation}
where $E_\pi\,,F_\pi$ are scalar functions independent of the relative momentum between the valence quark  and antiquark. As with the quark propagator, the simple structure of the BSA is a consequence of the CI and the corresponding symmetry-preserving regularization. The process for solving the pion BSE is described in Appendix~\ref{Bsekernel}. 

In order to compute physical observables, the obtained BSA must be canonically normalized. For the pseudoscalar case, the normalization condition reads
\begin{equation}\label{eqn::norm}
\!P_{\mu}=N_{c}tr\int_{q}\Gamma_{\pi}(-P)S(q)\Gamma_{\pi}(P)\frac{\partial}{\partial P_{\mu}}S(q-P)\,.
\end{equation}
The pion leptonic decay constant may be computed straightforwardly: this is expreassable in the following way:
\begin{equation}\label{eqn::decayd}
f_{\pi}P_{\mu}=N_{c}tr\int_q\gamma_5\gamma_{\mu}S(q)\Gamma_{\pi}(P)S(q-P)\,.
\end{equation}
Notably, our symmetry-preserving scheme ensures the Goldberger-Treiman relations are reproduced\,\cite{Gutierrez-Guerrero:2010waf}. In particular, in the chiral limit ($P^2=0=m$), we have
\begin{equation}\label{eqn::fpi}
    f_\pi E_\pi=M\,.
\end{equation}
The computed masses, decay constants and the normalized BS amplitudes of  the $\pi$ meson in the CI-MRL truncation, as well as the mass function of dressed quark, are reported in Table~\ref{tab:masscball}. It is important to highglight that these quantities are independent of the value of $\xi$ since the NL term in MRL do not contribute to the pseudoscalar meson BSE, see Appendix~\ref{Bsekernel}. Therefore, the static properties of the pion remains the same as those computed within the CI-RL case. This is by no means the case of the TFF, since it turns out that the NL pieces of the BS kernel influence the vector channels, so that the structure of the QPV changes favorably\,\cite{Xing:2021dwe}. This is discussed below.

\begin{table}[htpb]
\caption{\label{tab:masscball} Computed pion static properties. The model parameters: $m_G=0.132\,\GeV$, $\tau_{uv}=1/0.905\, \GeV^{-1}$ and  $\tau_{ir}=1/0.24\,\GeV^{-1}\simeq 1/\Lambda_{QCD}$. Mass units in $\GeV$. }
\setlength{\tabcolsep}{1.7mm}{
\begin{tabular}{c|ccccc}
\hline
$m$&$M$ &$m_{\pi}$  & $f_{\pi}$  &$E_{\pi}$ & $F_{\pi}$\\
\hline
0  &0.358 & 0 & 0.100 &3.566 &  0.458 \\
\hline
0.007&0.368 & 0.140 & 0.101 &3.595 &  0.475 \\
\hline
\end{tabular}}
\end{table}

\subsection{Quark-photon vertex}
The inhomogeneous BSE for the QPV $\Gamma_{\mu}(Q)$ in the MRL truncation is written as
\begin{eqnarray}\label{eqn::QPV}
    \Gamma_{\mu}(Q)=\gamma_\mu&-&\frac{4}{3m_G^2}\int_q \gamma_\alpha S(q)\Gamma_{\mu}(Q)S(q-Q)\gamma_\alpha\nn
    &+&\frac{4\xi}{3m_G^2}\int_q \tilde{\Gamma}_j S(q)\Gamma_{\mu}(Q)S(q-Q)\tilde{\Gamma}_j.
\end{eqnarray}
The simplicity of the CI model enable us to fully characterize the QPV by three tensor structures, namely
\begin{equation}
 \label{eq:QPV}
    \Gamma_{\mu}(Q)=\gamma^L_\mu f_L(Q^2)+\gamma^T_\mu f_T(Q^2)+\frac{\sigma_{\mu\nu}Q_\nu}{M} f_A(Q^2)\,,
\end{equation}
where $\gamma_\mu^T=\gamma_\mu-\frac{\slashed{Q}Q_\mu}{Q^2}$, $\gamma_\mu^L=\gamma_\mu-\gamma_\mu^T$. By solving \Eqn{eqn::QPV}, one obtains the dressing functions $f_L(Q^2)$, $f_T(Q^2)$ and $f_A(Q^2)$:
\begin{eqnarray}
\label{eq:QPVdress}
    f_L(Q^2)&=&1\,,\nn
    f_T(Q^2)&=&-\frac{\mathcal{I}}{\mathcal{K}(C_0^2 M^2 \hat{\xi}+2 \Bar{C}_0\mathcal{I})-\mathcal{I}}\,,\nn
    f_A(Q^2)&=&\frac{C_0 M^2 \hat{\xi}}{\mathcal{K}(C_0^2 M^2 \hat{\xi}+2 \Bar{C}_0\mathcal{I})-\mathcal{I}}\,;
\end{eqnarray}
the integrals $C_\alpha, \Bar{C}_\alpha$ are defined as follows:
\begin{eqnarray}
    C_\alpha(Q^2)&=&\int_0^1 I_\alpha(\omega(M^2,u,Q^2)) du,\nn
    \Bar{C}_\alpha(Q^2)&=&\int_0^1 u(u-1) I_\alpha(\omega(M^2,u,Q^2)) du\,,
\end{eqnarray}
with $\omega=M^2+u(1-u)Q^2$, $\mathcal{I}=1-\hat{\xi}(2 C_0 M^2+C_2)$, $\hat{\xi}=\frac{32\xi}{9m_G^2}$ and $\mathcal{K}=\frac{8 Q^2}{3 m_G^2}$.

 Let us now analyze the QPV dressing functions from Eq.\,\eqref{eq:QPVdress}. Firstly, the longitudinal piece $f_L(Q^2)$ ensures the vertex satisfies the symmetry requirement of the Ward-Green-Takahashi identity\,
 \cite{Albino:2018ncl}. Second, the transverse dressing function $f_T(Q^2)$ exhibits a vector meson pole in the timelike axis. This would also be the case for RL truncation\,\cite{Maris:1999bh}, although the mass of the vector meson would be shifted due to the influence of the NL pieces in the BS kernel\,\cite{Xing:2021dwe}.  Notably, the dressing function of the AMM term, $f_A(Q^2)$, happens to be proportional to the strength parameter $\xi$, so that this term vanishes in the $\xi = 0$ limit. It is precisely in this case that one recovers the RL truncation, which indicates that within the CI, the RL is unable to generate an AMM piece naturally; in fact, such term is often added by hand\,\cite{Wilson:2011aa,Raya:2021pyr}. Regarding the asymptotic falloff of the dressing functions, one observes that $f_T(Q^2\to \infty) \to 1$ and $f_A(Q^2 \to \infty) \to 0$, so Eqs.\,(\ref{eq:QPV}, \ref{eq:QPVdress}) would guarantee that the tree-level result, $\gamma_\mu$, is faithfully recovered. Finally, it is worth showing the values of these dressing functions in the $Q^2=0$ point:
\begin{eqnarray}
    f_L(0)&=&1,\nn
    f_T(0)&=&1,\nn
    f_A(0)&=&\frac{\hat{\xi}M^2I_0(M^2)}{1-\hat{\xi}(2M^2I_0(M^2)+I_2(M^2))}\,.
\end{eqnarray}
Thus, as we shall discuss in the upcoming section, the AMM could indeed produce a quantifiable contribution to the two-photon TFF, $\gamma^* \pi^0 \gamma$. Some illustrations of the QPV dressing functions are found in Ref.\,\cite{Xing:2021dwe}.

\section{\texorpdfstring{$\gamma^{*}\pi^0\gamma$}{} process and the chiral anomaly}\label{sec::ano}
Let us now focus on the $\gamma^{*}\pi^0\gamma$ transition process, which is parametrized by the matrix element (with $e$ being the unit charge):
\begin{equation}
    T_{\mu\nu}(k_1,k_2)=\frac{e^2}{4\pi^2f_\pi}\epsilon_{\mu\nu k_1 k_2}G(k_1^2,k_1 \cdot k_2,k_2^2)\,;
\end{equation}
here $\epsilon_{\mu\nu k_1 k_2}=\epsilon_{\mu\nu \alpha \beta}k_{1\alpha}k_{2\beta}$ is understood. In the impuse approximation, this transition is expressed as\,\cite{Maris:2002mz,Raya:2015gva,Chen:2016bpj}:
\begin{eqnarray}\label{eqn::transition}
    T_{\mu\nu}(k_1,k_2)&=&\text{tr}\int_qi\Gamma_{\nu}(k_2)S(q-k_2)\nn
    &\times&\Gamma_{\pi}(P)S(q+k_1)i\Gamma_{\mu}(k_1)S(q),
\end{eqnarray}
where the trace \text{tr} is taken over the Dirac indices and $P=-(k_1+k_2)$ is the pion's total momentum, such that $P^2=-m_\pi^2$. If $k_1$ denotes the momentum of the off-shell photon, then the corresponding kinematic constraints read
\begin{equation}\label{kcon}
    k_1^2=Q^2,\ k_2^2=0,\ k_1\cdot k_2=-(Q^2+m_\pi^2)/2.
\end{equation}
By adopting these kinematics, the $G^{\gamma^{*}\pi^0\gamma}(Q^2)$ TFF is then defined as: 
\begin{eqnarray}\label{eqn::transitionff}
   G^{\gamma^{*}\pi^0\gamma}(Q^2)=2G(Q^2,-(Q^2+m_\pi^2)/2,0),
\end{eqnarray} 
where the factor $2$ appears in order to account for the possible ordering of the photons.

With all the elements entering Eq.\,\eqref{eqn::transition}, namely quark propagator,  pion BSA and quark photon vertex determined in Sec.~\ref{sec:ci}, it is in principle straightforward to compute this transition. 
However, certain ambiguities caused by the definition of the $\gamma_5$ matrix arise in treatments that require a regularization scheme. In particular, the trace involving odd numbers of $\gamma_5$ matrices leads to different results, as can be seen in Eqs.~(3.11, 3.13. 3.16) from Ref.~\cite{Ma:2005md}. Although it can be proved that these different results can be transformed into one another, the adopted definition of $\gamma_5$ might influence the final outcomes, and so is the case the chiral anomaly. 
To overcome these issues, Ref.~\cite{Ma:2005md} suggests the following definition of $\gamma_5$
\begin{equation}
    \gamma_5=-\frac{1}{24}\epsilon_{abcd}\gamma_{a}\gamma_{b}\gamma_{c}\gamma_{d}
\end{equation}
to evaluate traces that contains odd number of $\gamma_5$, so we will adopt this choice in the subsequent.

Let's first focus on the chiral anomaly, which is associated with on-shell photon ($Q^2=0$) and chiral limit pion ($m_\pi = 0$), \emph{i.e.}, $G(0,0,0)$. The computed result might be represented as 
\begin{equation}
    \frac{1}{4\pi^2f_\pi}G(0,0,0)=\frac{4N_c}{3M}\left(G_{E}(0,0,0)E_\pi+G_{F}(0,0,0)F_\pi\right),
\end{equation}
where
\begin{eqnarray}
   G_{E}(0,0,0)&=& M^2 I_{-2}\left(M^2\right) f_T^2(0)\label{eq:GE}\\
   &+&\left[I_0\left(M^2\right)+4 M^2 I_{-2}\left(M^2\right)\right]f_T(0)f_A(0)\nn
   &+&\left[I_0\left(M^2\right)+4 M^2 I_{-2}\left(M^2\right)\right]f_A^2(0),\nonumber\\
   G_{F}(0,0,0)&=&0.
\end{eqnarray}
The first thing to note is that the pseudovector component of the pion BSA, $F_\pi$, does not contribute the anomaly. This has been shown to be the case for any symmetry-preserving treatment, based upon DSE and BSEs, of the pion TFF\,\cite{Maris:1998hc}. By employing the regularization scheme introduced in Ref.~\cite{Xing:2022jtt}, the present computation of the pion TFF falls upon this category. 

\begin{figure}[hptb]
\includegraphics[width=8.6cm]{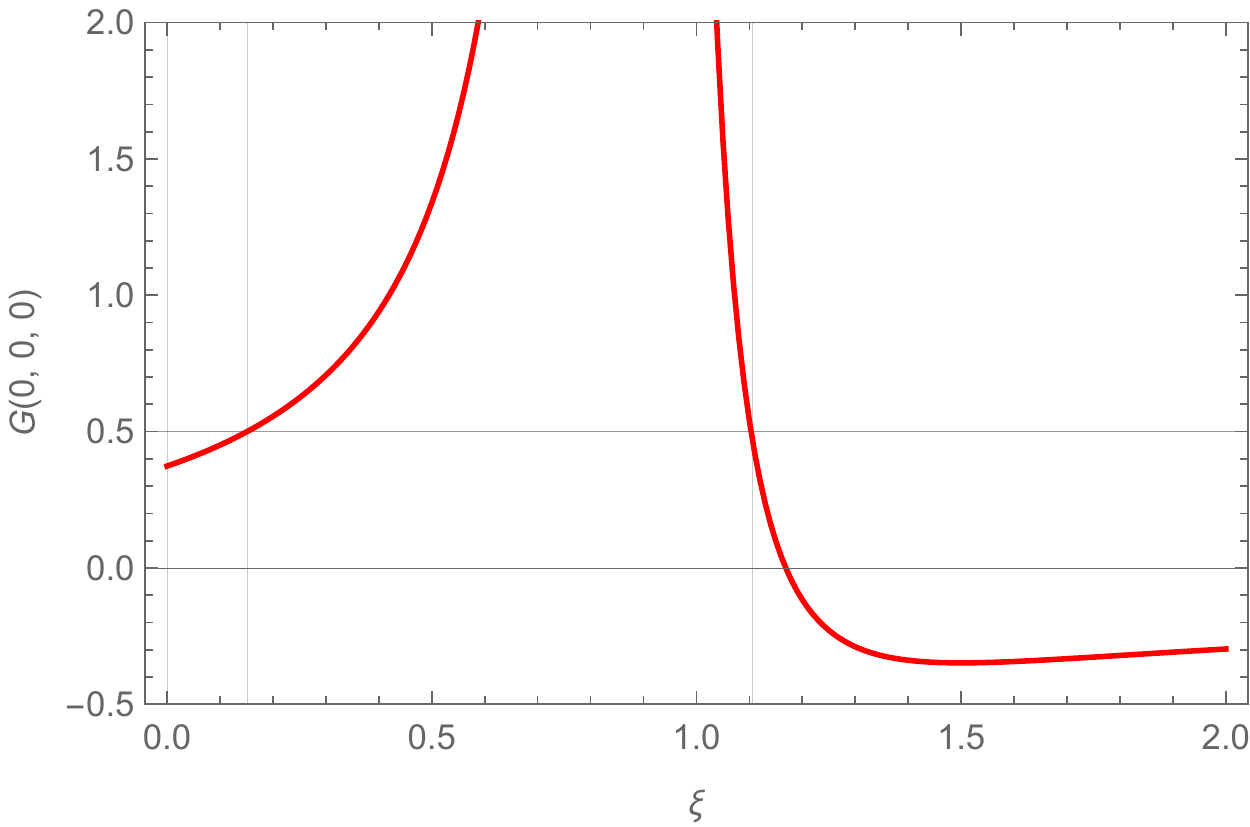}
\caption{The variation of $G(0,0,0)$ as a function of the AMM strength parameter $\xi$.\label{xi}}
\end{figure}

Focusing on the non-vanishing contribution, $G_E$, the last two lines in Eq.\,\eqref{eq:GE} reveal that the AMM term in the QPV indeed contribute the chiral anomaly. To address this observation, let us consider the $\xi=0$ limit, corresponding to the CI-RL result~\cite{Roberts:2010rn}:
\begin{equation}
    G^{\xi=0}(0,0,0)=16 \pi^2 M^2 I_{-2}\left(M^2\right)\,,
\end{equation}
where we have employed Eq.\,\eqref{eqn::fpi}. Since $I_{-2}\left(M^2\right)$ is convergent, one can take the limits $\tau_{uv}\to 0$, $\tau_{ir}\to\infty$, and find out that
\begin{equation}
       \left.I_{-2}(M^2)\right|_{\tau_{uv}\to 0}^{\tau_{ir}\to\infty}=\frac{1}{16\pi^2}\frac{1}{2M^{2}}\,.
\end{equation}
Thus, it turns out that the famous chiral anomaly is  recovered by removing the cutoffs:
\begin{equation}
    G^{\xi=0}(0,0,0)|_{\tau_{uv}\to 0}^{\tau_{ir}\to\infty}=\frac{1}{2}\,.
\end{equation}
Thus, as we have seen, the chiral anomaly would only be recovered after the cutoffs are eliminated. However, in an effective theory, the ultraviolet cutoff is part of the theory itself and, in principle, must not be removed. Under such circumstances, $G^{\xi=0}(0,0,0)\neq \frac{1}{2}$, thus failing at meeting the value dictated by the chiral anomaly. According to Ref.~\cite{Alkofer:1992nh}, higher order contributions are responsible for the missing part of the anomaly. The analysis of Ref.~\cite{Lepage:1997cs} shows how sensible terms can be added to the theory in order to mimic the complex short-distance dynamics left out by the momentum cutoff. However, for these additional terms to be given this interpretation, it is necessary for their contribution to vanish when the UV cutoff is removed. Here is where the quark AMM term in the QPV, generated automatically by the MRL truncation, becomes crucial:  firstly, according to Eq.\,\eqref{eq:GE}, it provides a non-vanishing contribution to $G(0,0,0)$; secondly, for arbitrary $\xi$ its contribution vanishes when the cutoffs are removed:
\begin{equation}
    G(0,0,0)^{\xi \neq 0}|_{\tau_{uv}\to 0}^{\tau_{ir}\to\infty}=\frac{1}{2}\,,
\end{equation}
\emph{i.e.} it can be duly identified as an effective term that properly encodes the short-distance dynamics beyond the cutoff. Thus, we can readily fix the strength parameter $\xi$ by requiring $G(0,0,0)=\frac{1}{2}$ with finite cutoffs. The variation of $G(0,0,0)$ with $\xi$ is depicted in \Fig{xi}. It is seen that $G(0,0,0)=\frac{1}{2}$ is in fact obtained from two possible values of $\xi$. Nonetheless, the largest one is unphysical since the $\rho$ meson BSE yields to no bound state solutions in this case. We therefore adopt the smallest value $\xi=0.151$ and employ it to evaluate the two-photon TFF.

To investigate the contributions of the difference pieces of the QPV on the $\gamma^* \pi^0 \gamma$ TFF, the latter is computed in three different cases: by employing the \emph{complete} QPV derived from Eq.\,\eqref{eq:QPV} (with our preferred value $\xi = 0.151$), by taking the CI-RL truncation limit (corresponding to $\xi=0$), thus neglecting the quark AMM contribution, and by simply pluggin in the tree-level form of the the QPV, thus discharging the AMM term and vector meson pole contributions. The resulting form factors are shown in Fig.\,\ref{GQ2}, and the explicit mathematical expressions are presented in Appendix~\ref{formfactor}. Clearly, having retained the cutoffs, the second and third cases (RL and tree level vertices, respectively) fail to obtain the correct normalization of the form factor. This is not the case when the quark AMM is properly incorporated; not only the correct normalization is obtained (thus reproducing the chiral anomaly), but the TFF exhibits the steepest falloff among the three cases, becoming practically indistinguishable from the realistic QCD-based computations\,\cite{Maris:2002mz,Raya:2015gva,Chen:2016bpj} in the small $Q^2$ domain.  Conversely, it is conspicuously visible the influence of the dressing functions becomes more irrelevant with increasing photon virtuality, and the hardness of the TFFs prevails in any case. This outcome is expected due to the momentum-independent nature of the CI model,\,\cite{Gutierrez-Guerrero:2010waf,Roberts:2010rn}.

At the $Q^2 \to 0$ limit, in the MRL case, we also compute the neutral pion decay width and the corresponding interaction-radius. These are defined, respectively, as follows: 
\cite{Maris:2002mz}
\begin{eqnarray}\label{eqn::decay}    \Gamma_{\pi^0\gamma\gamma}&=&\frac{g^2_{\pi\gamma\gamma}(Q^2)\alpha^2_{em} m^3_{\pi}}{16\pi^3 f^2_\pi}|_{Q^2=0}\,\\
\label{eqn::radius}
    r_{\pi^0}^2&=&-6\frac{d}{dQ^2} \ln g_{\pi\gamma\gamma}(Q^2)|_{Q^2=0},
\end{eqnarray}
where $g_{\pi\gamma\gamma}(Q^2)=G^{\xi=0.151}(Q^2,-(Q^2+m_\pi^2)/2,0)$ and $\alpha_{em}=1/137$. \Eqn{eqn::decay} yields $\Gamma=7.21 \,\eV$, in agreement with the experimental determination $\Gamma=7.82\pm0.14\pm0.17\,\eV$\,\cite{ParticleDataGroup:2022pth}. This compatibility is not surprising since, by satisfying the anomaly in the chiral limit, the prediction at the physical pion mass becomes practically independent of the model inputs. Furthermore, the interaction-radius computed from \Eqn{eqn::radius} is $r_{\pi^0}=0.61\,\fm$, which is also in fair agreement with the experimental estimate $r_{\pi^0}=0.65\pm0.03\, \fm$\,\cite{cello1991measurement}. The presence of the quark AMM is crucial in this case, otherwise the produced value of $r_{\pi^0}$ would be practically halved\,\cite{Roberts:2010rn}.



\begin{figure}[hptb]
\includegraphics[width=8.6cm]{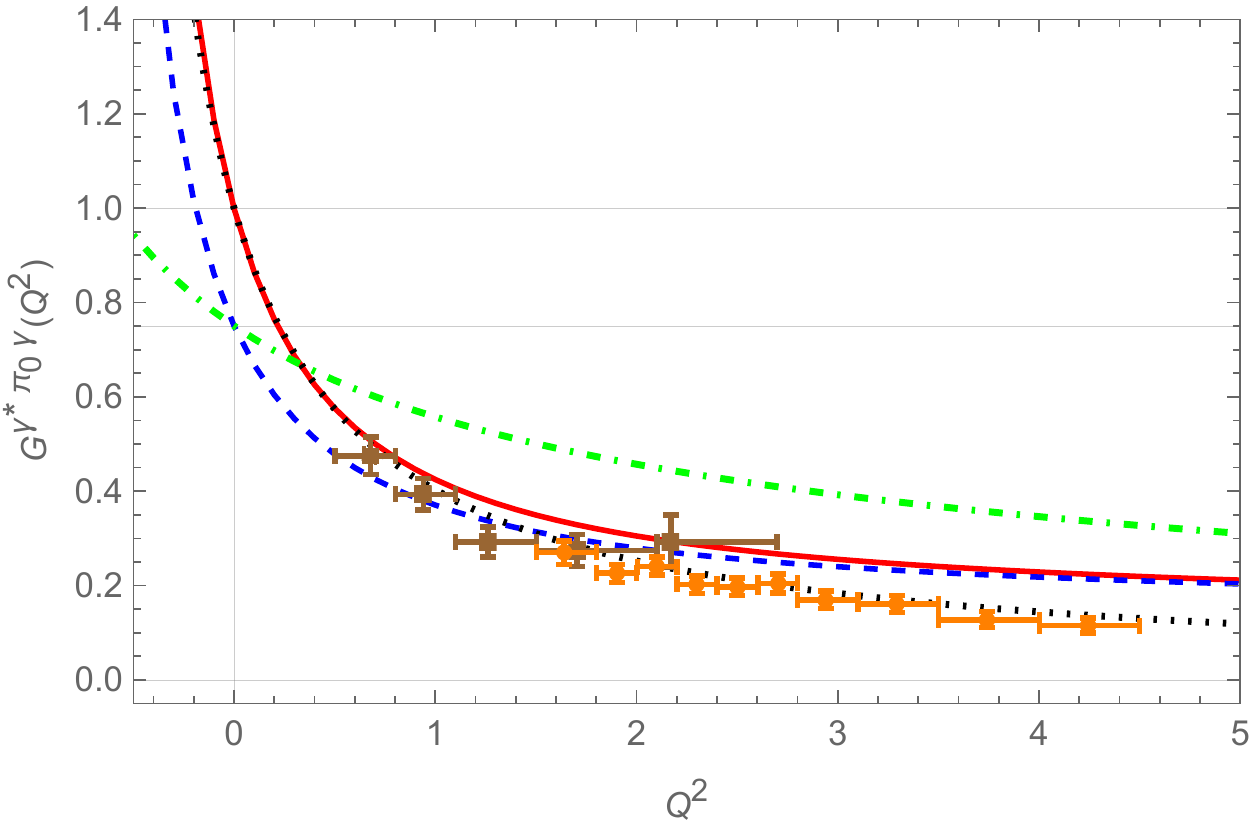}
\caption{$\gamma^{*}\pi^0\gamma$ transition form factor. Solid curve-full computation with MRL truncation; dashed-results calculated with RL truncation; dash-dotted- results obtained with bare QPV. The dotted curve corresponds to a monopole ﬁt to the QCD-based result in Ref.\,\cite{Maris:2002mz}. Experimental data from Refs. \cite{cello1991measurement,gronberg1998measurements}, Brown polygons and Orange disks, respectively. Mass units in GeV.\label{GQ2}}
\end{figure}

\section{Conclusion}\label{sec::con}
In this work we have computed the $\gamma^{*}\pi^0\gamma$ TFF, intimately connected with the so called chiral anomaly in the Standard Model, as well as the associated decay width and interaction radius. The calculation is based upon a symmetry-preserving model of QCD, embedded within the so called MRL truncation. In addition to the soundness of the RL approximation to address static and structural properties of pseudoscalar mesons, the MRL scheme enables the CI to produce a quark AMM term in the QPV. While RL and MRL would produce the same static properties of the pion, Table~\ref{tab:masscball}, the presence of the quark AMM would be crucial in the evaluation of the two-photon TFF and, in particular, in correctly reproducing the chiral anomaly.

Let us now recall that in effective field theories, the chiral anomaly might be compromised due to the presence of finite cutoffs\,\cite{Alkofer:1992nh,Lepage:1997cs}, whose presence could neglect complex dynamics that otherwise would have an effect on the anomaly. It is argued that such higher-order effects can be enconded in additional terms that must vanish when the cutoffs are removed\,\cite{Lepage:1997cs}. This is the case of the quark AMM. To support this statement, firstly note that in the computation of the TFF only the leading component of the pion BSA, $E_\pi$, contributes to the anomaly, \emph{i.e.} $G_F(0,0,0)=0$. This is nothing but a consequence of the symmetry-preserving regularization treatment of the CI described herein. On the other hand, the quark AMM does indeed contribute to the normalization of the form factor; thereby, the strength parameter $\xi$ can be tuned to produce, $G(0,0,0)=1/2$, as imposed by the anomaly. Furthermore, when the cutoffs are removed, the contribution arising from the AMM vanishes regardless of the value of $\xi$. This is sufficient to adopt for this piece the interpretation of Ref.\,\cite{Lepage:1997cs}:  the AMM term could be adequately regarded as an effective term that simulates the physics discarded by the cutoffs. This interpretation becomes even more natural due to the fact that both chiral anomaly and quark AMM are strongly influenced by the effects of DCSB\,\cite{Roberts:1994hh,Chang:2010hb,Bashir:2011dp}.

Our numerical evaluation of the $\gamma^* \pi^0 \gamma$ transition included three different inputs for the QPV: the fully-dressed QPV obtained in connection with the CI-MRL truncation, the one derived in the CI-RL approximation (that neglects the AMM piece), and the tree level vertex. Firstly, let us note that the last two cases fail to reproduce the correct normalization of the form factor; it is only possible to obtain if the cutoffs are removed\,\cite{Roberts:2010rn}. The CI-MRL computation, on the other hand, satifies the chiral anomaly while also being practically indistinguishable from the QCD-based results,\,\cite{Maris:2002mz,Raya:2015gva,Chen:2016bpj}, at low $Q^2$. In this case, the values obtained for the decay widths and interaction radius are found to be compatible with the empirical determinations as well. As the virtuality of the photon grows, the QPV dressing functions cease to be relevant and the three cases are reduced to the same; in this domain of photon momentum, the form factors become harder, as one would expect from the CI model. Finally, it is important to highlight that such encouraging results for the CI-MRL case indicate the effectiveness of the idea implemented in this work in dealing with chiral anomaly in effective theories.

This is an encouraging step towards a comprehensive study of hadrons in CI-MRL approach. Immediate next step will involve to calculate $\gamma\pi^{\star}\to\pi\pi$ TFF, and study the chiral anomaly in this process. The work on $\gamma\pi^{\star}\to\pi\pi$ TFF is underway.


\hspace*{\fill}\ 
\begin{acknowledgments}
L. Chang is grateful for constructive conversations with Li-Sheng Geng and Mao-Zhi Yang. Work supported by National Natural Science Foundation of China (grant no. 12135007).
\end{acknowledgments}
\appendix\widetext
\section{}\label{Bsekernel}

Inserting \Eqn{eqn::pibsa} into \Eqn{eqn::pion}, the general form of the pion BSA into the corresponding BSE, one obtains the following coupled equations for the scalar functions $E_\pi$ and $F_\pi$:
\begin{equation}\label{eqn::pionm}
\begin{aligned}
\left[\begin{array}{c}E_\pi\\F_\pi\end{array}\right]=\frac{4}{3 m_G^2}\left[\begin{array}{cc}K_{EE}&K_{EF}\\K_{FE}&K_{FF}\end{array}\right]\left[\begin{array}{c}E_\pi\\F_\pi\end{array}\right].
\end{aligned}
\end{equation}
By taking Dirac trace and following the aforementioned regularization procedure, the kernels can be written as (with $\omega=M^2+u(1-u)P^2$):
\begin{eqnarray}
    K_{EE}&=&4\int_{0}^{1}du I_2(\omega)-2u(1-u)P^2I_0(\omega)\nn
    K_{EF}&=&4\int_{0}^{1}du P^2I_0(\omega)\nn
    K_{FE}&=&2\int_{0}^{1}du M^2I_0(\omega)\nn
    K_{FF}&=&-4\int_{0}^{1}du M^2I_0(\omega).
\end{eqnarray}
Solutions to the eigenvalue equation, \Eqn{eqn::pionm}, are only found at discrete values of momentum $P_i$. Thus, it is convenient to introduce the eigenvalue $\lambda(P^2)$ to deal with this equation, namely:
 \begin{equation}
\begin{aligned}
\lambda(P^2)\left[\begin{array}{c}E_\pi\\F_\pi\end{array}\right]=\frac{4}{3 m_G^2}\left[\begin{array}{cc}K_{EE}&K_{EF}\\K_{FE}&K_{FF}\end{array}\right]\left[\begin{array}{c}E_\pi\\F_\pi\end{array}\right],
\end{aligned}
\end{equation}
The smallest value $P_i^2$ producing $\lambda(P_i^2) = 1$, corresponds to the ground-state pion, such that $P^2=-m_\pi^2$.

\section{}\label{formfactor}
With the kinematic constraints \Eqn{kcon}, the transition form factor can be written as
\begin{equation}
  \frac{1}{4 \pi^2 f_\pi}G(Q^2,-(Q^2+m_\pi^2)/2,0)=\frac{4N_c}{3M}\int_0^1 du_1\int_0^{1-u_1}du_2(G_E(Q^2,u_1,u_2) E_\pi +G_F(Q^2,u_1,u_2) F_\pi),
\end{equation}
where, with $\rho(M,u_1,u_2,Q^2)=M^2+u_1(1-u_1)Q^2-u_1u_2(Q^2+m_\pi^2)$, one gets:
\begin{equation}\label{b1}
    \begin{aligned}
         G_E(Q^2,u_1,u_2)=&2I_{-2}(\rho)M^2 f_T^2\\
         +&(2I_{-2}(\rho)u_1 Q^2-6I_{-2}(\rho)u_1^2 Q^2+4I_{-2}(\rho)u_1^3Q^2+8I_{-2}(\rho)u_1^2u_2Q^2-6I_{-2}(\rho)u_1u_2Q^2\\
         +&4I_{-2}(\rho)u_1u_2^2Q^2+3I_0u_1+3I_0u_2+8I_{-2}(\rho)M^2+4I_{-2}(\rho)u_1^2u_2m_\pi^2-4I_{-2}(\rho)u_1u_2m_\pi^2\\
         +&4I_{-2}(\rho)u_1u_2^2m_\pi^2)f_A f_T\\
         +&(2I_{-2}u_1Q^2-4I_{-2}u_1^2Q^2-4I_{-2}u_1u_2Q^2+2I_0+8I_{-2}M^2-4I_{-2}u_1u_2m_\pi^2)f_A^2\\
         G_F(Q^2,u_1,u_2)=&(6I_{-2}(\rho)u_1Q^2-10I_{-2}(\rho)u_1^2Q^2+4I_{-2}(\rho)u_1^3Q^2+8I_{-2}(\rho)u_1^2u_2Q^2-10I_{-2}(\rho)u_1u_2Q^2\\
         +&4I_{-2}(\rho)u_1u_2^2Q^2-2I_{0}(\rho)+3I_{0}(\rho)u_1+3I_{0}(\rho)u_2+4I_{-2}(\rho)u_1^2u_2m_\pi^2-8I_{-2}(\rho)u_1u_2m_\pi^2\\
         +&4I_{-2}(\rho)u_1u_2^2m_\pi^2 )f_T^2\\
         +&(2I_{-2}(\rho)Q^2+I_{-2}(\rho)u_1Q^2-8I_{-2}(\rho)u_1^2Q^2-8I_{-2}(\rho)u_1u_2Q^2-2I_{-2}(\rho)u_2Q^2\\
         -&2I_{-2}(\rho)u_1m_\pi^2-8I_{-2}(\rho)u_1u_2m_\pi^2-2I_{-2}(\rho)u_2m_\pi^2)f_Tf_A\\
         +&(-I_{0}(\rho)u_1\frac{Q^2}{M^2}+4I_{-2}(\rho)Q^2+2I_{-2}(\rho)u_1\frac{Q^4}{M^2}-4I_{-2}(\rho)u_1Q^2-6I_{-2}(\rho)u_1^2\frac{Q^4}{M^2}\\
         +&4I_{-2}(\rho)u_1^3\frac{Q^4}{M^2}+8I_{-2}(\rho)u_1^2u_2m_\pi^2\frac{Q^2}{M^2}+6I_{-2}(\rho)u_1^2u_2\frac{Q^4}{M^2}-4I_{-2}(\rho)u_1u_2m_\pi^2\frac{Q^2}{M^2}\\
         -&4I_{-2}(\rho)u_1u_2\frac{Q^4}{M^2}+4I_{-2}(\rho)m_\pi^2\frac{Q^2}{M^2}+2I_{-2}(\rho)u_1u_2^2\frac{Q^4}{M^2}-4I_{-2}(\rho)u_2Q^2-4I_{-2}(\rho)u_1m_\pi^2\\
         +&2I_{-2}(\rho)u_1^2u_2\frac{m_\pi^4}{M^2}+2I_{-2}(\rho)u_1u_2^2\frac{m_\pi^4}{M^2}-4I_{-2}(\rho)u_2m_\pi^2)f_A^2
    \end{aligned}
\end{equation}
\bibliography{bibreferences}
\end{document}